\documentclass[multphys,vecphys]{svmult}

\usepackage{makeidx}         
\usepackage{multicol}        
\usepackage[bottom]{footmisc}

\makeindex             
\def\eth{\epsilon}
\def\emx{\varepsilon}
\def\ER{E_{\rm R}}
\def\SD{{\rm SD}}
\def\SI{{\rm SI}}

\begin{document}

\title*{On the mixed spin-scalar coupling approach in dark matter search}

\author{V.A.Bednyakov\inst{1} 
        \and H.V.~Klapdor-Kleingrothaus\inst{2}}
\institute{Laboratory of Nuclear Problems,
         Joint Institute for Nuclear Research, 141980 Dubna, Russia. \
         \texttt{E-mail: Vadim.Bednyakov@jinr.ru}
\and     Max-Planck-Institut f\"{u}r Kernphysik, 
         Postfach 103980, D-69029, Heidelberg, Germany. \
         \texttt{E-mail: H.Klapdor@mpi-hd.mpg.de}}
\maketitle

{\bf Abstract.} 
        To avoid misleading discrepancies between results of 
        different dark matter search experiments 
        as well as between the data and SUSY calculations it is 
        in general
        preferable to use a mixed spin-scalar coupling approach in which 
        spin-independent and spin-dependent WIMP-nucleon
        couplings are {\em both}\/ non-zero.  
        On the other hand one may, however, 
        to safely neglect the
        subdominant spin WIMP-nucleon contribution
        in comparison with the spin-independent one 
        in analysis of 
        data from experiments with 
        heavy enough non-zero-spin target nuclei.

        The mixed coupling approach is applied to estimate 
        future prospects of experiments with the odd-neutron 
        high-spin isotope $^{73}$Ge.

\section{Introduction}
        In many experiments one tries 
        to detect directly 
        relic dark matter (DM)
        Weakly Interacting Massive Particles (WIMPs) $\chi$ via 
        their elastic scattering on a target nucleus $(A,Z)$. 
        The nuclear recoil energy 
        $E_{\rm R}$ ($E_{\rm R}\sim 10^{-6} m_{\chi} \sim$ few keV) 
        is measured. 
        The expected differential event rate has the form 
\cite{Jungman:1996df}--\cite{Bednyakov:1994qa}:
\begin{eqnarray}
\label{Definitions.diff.rate}
\frac{dR}{dE_{\rm R}} &=& N_t \frac{\rho_\chi}{m_\chi}
        \displaystyle \int^{v_{\max}}_{v_{\min}} dv f(v) v
        {\frac{d\sigma}{dq^2}} (v, q^2),
        \qquad E_{\rm R} = q^2 /(2 M_A ). 
\end{eqnarray}
        Here, 
        $v_{\min}=\sqrt{M_A E_{\rm R}/2 \mu_{A}^2}$, 
        $v_{\max} = v_{\rm esc} \approx 600$~{km}/{s}, 
        $\mu_A = \frac{m_\chi M_A}{m_\chi+ M_A}$; 
        $f(v)$ is the distribution of 
        $\chi$-particles in the solar vicinity, 
        $N_t$ 
        is the number density of target nuclei. 
        $M_A$ denotes the target nuclear mass, 
        the dark matter density is usually assumed to be
        $\rho_{\chi}$ = 0.3 GeV$/$cm$^{3}$. 
        The $\chi$-nucleus elastic scattering cross section 
        for non-zero-spin ($J\neq 0$) nuclei is a sum of
        the spin-independent (SI, or scalar) 
        and spin-dependent (SD, axial) terms
\cite{Engel:1991wq,Engel:1992bf,Ressell:1993qm,Ressell:1997kx}:
\begin{eqnarray} 
{\frac{d\sigma^{A}}{dq^2}}(v,q^2) 
\label{Definitions.cross.section}
        &=& \frac{\sigma^A_{\rm SD}(0)}{4\mu_A^2 v^2}F^2_{\rm SD}(q^2)
           +\frac{\sigma^A_{\rm SI}(0)}{4\mu_A^2 v^2}F^2_{\rm SI}(q^2).
\end{eqnarray}
        For $q=0$ the nuclear SD and SI cross sections take the forms  
\begin{eqnarray}
\label{NuclCS0}
\label{Definitions.scalar.zero.momentum}
\sigma^A_{\rm SI}(0) 
        &=& \frac{\mu_A^2}{\mu^2_p}A^2 \sigma^{p}_{{\rm SI}}(0), \\ 
\label{Definitions.spin.zero.momentum}
\sigma^A_{\rm SD}(0)
        &=& \frac{4\mu_A^2}{\pi}\frac{(J+1)}{J}
             \left\{a_p\langle {\bf S}^A_p\rangle 
                  + a_n\langle {\bf S}^A_n\rangle\right\}^2 \\
&=&
        \frac{\mu_A^2}{\mu_p^2}\frac43 \frac{J+1}{J}
        \sigma^{}_{\rm SD}(0)   
        \left\{ \langle {\bf S}^A_p\rangle \cos\theta
               +\langle {\bf S}^A_n\rangle \sin\theta   
       \right\}^2.
\label{Definitions.spin.zero.momentum.Bernabei}
\end{eqnarray}
        The dependence on effective $\chi$-quark 
        scalar ${\cal C}_{q}$ and axial ${\cal A}_{q}$ couplings 
        and on the spin $\Delta^{(p,n)}_q$
        and the mass $f^{(p,n)}_q$ structure of {\em nucleons}\ 
        enter into these formulas via the zero-momentum-transfer 
        proton and neutron SI and SD cross sections
        ($\mu^2_{n}=\mu^2_{p}$ is assumed): 
\begin{eqnarray}
\sigma^{p}_{{\rm SI}}(0) 
        = 4 \frac{\mu_p^2}{\pi}c_{0}^2,
&\qquad&
\sigma^{p,n}_{{\rm SD}}(0)  
        =  12 \frac{\mu_{p,n}^2}{\pi}{a}^2_{p,n}; \\
        c_{0} = c^{p,n}_0 = \sum_q {\cal C}_{q} f^{(p,n)}_q,
&\quad&
        a_p =\sum_q {\cal A}_{q} \Delta^{(p)}_q, \quad 
        a_n =\sum_q {\cal A}_{q} \Delta^{(n)}_q.
\label{a_pn}
\end{eqnarray}
        The effective spin WIMP-nucleon cross section
        $\sigma^{}_{\rm SD}(0)$
        and the coupling mixing angle $\theta$ 
\cite{Bernabei:2003za,Bernabei:2001ve} were introduced in 
(\ref{Definitions.spin.zero.momentum.Bernabei}):
\begin{eqnarray}
\label{effectiveSD-cs}
\sigma^{}_{\rm SD}(0)
        &=& \frac{\mu_p^2}{\pi}\frac43 
                \Bigl[ a_p^2 +a_n^2 \Bigr], \qquad
\tan\theta = \frac{{a}_{n}}{{a}_{p}}. 
\end{eqnarray}
        The factors $\Delta_{q}^{(p,n)}$, which parametrize the quark 
        spin content of the nucleon, are defined as
        $ \displaystyle 2 \Delta_q^{(n,p)} s^\mu \equiv 
          \langle p,s| \bar{\psi}_q\gamma^\mu \gamma_5 \psi_q    
          |p,s \rangle_{(p,n)}$.
        The $\langle {\bf S}^A_{p(n)} \rangle$ is the total 
        spin of protons 
        (neutrons) averaged over all $A$ nucleons of the nucleus $(A,Z)$.
        In the simplest case the SD and SI nuclear form-factors
\begin{equation}
\label{Definitions.form.factors}
F^2_{\rm SD,SI}(q^2) = \frac{S^{A}_{\rm SD,SI}(q^2)}{S^{A}_{\rm SD,SI}(0)}
\end{equation}
        have a Gaussian form (see, for example, 
\cite{Ellis:1988sh}).
        The spin-dependent structure function 
        $S^A_{\rm SD}(q)$ 
        in terms of isoscalar $a_0 = a_n + a_p$ and isovector 
        $a_1 = a_p - a_n$ effective couplings has the form
\cite{Ressell:1993qm,Ressell:1997kx}:
\begin{equation}
S^A_{\rm SD}(q) = a_0^2 S_{00}(q) + a_1^2 S_{11}(q) + a_0 a_1 S_{01}(q).
\label{Definitions.spin.decomposition} 
\end{equation}

\section{One-coupling dominance approach} 
\begin{figure}[ht!] 
\begin{picture}(100,115)
\put(-2,-5){\includegraphics{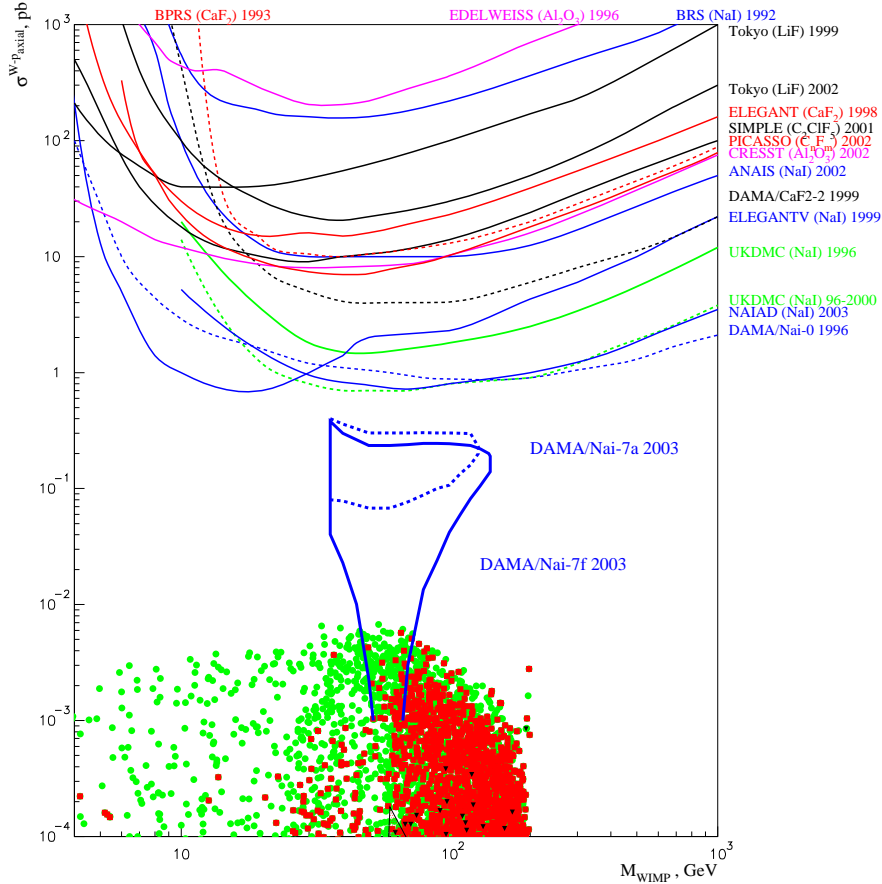}}
\end{picture}
\caption{Exclusion curves for the spin-dependent WIMP-proton cross sections
        ($\sigma^{p}_{{\rm SD}}$ as a function of the WIMP mass).
        All curves, except the NAIAD and Tokio-LiF, are 
        obtained in the one-coupling dominance approach 
        with $\sigma^{}_{\rm SI}=0$ and $\sigma^{n}_{{\rm SD}}=0$.
        DAMA/NaI-7a(f) contours for the WIMP-proton SD interaction in $^{127}$I
        are obtained on the basis of the positive 
        signature of annual modulation within the 
        framework of the mixed scalar-spin coupling approach
\cite{Bernabei:2003za,Bernabei:2001ve}. 
From \cite{Bednyakov:2004be}.}
\label{Spin-p} 
\end{figure}

\begin{figure}[ht] 
\begin{picture}(100,90)
\put(0,-4){\includegraphics{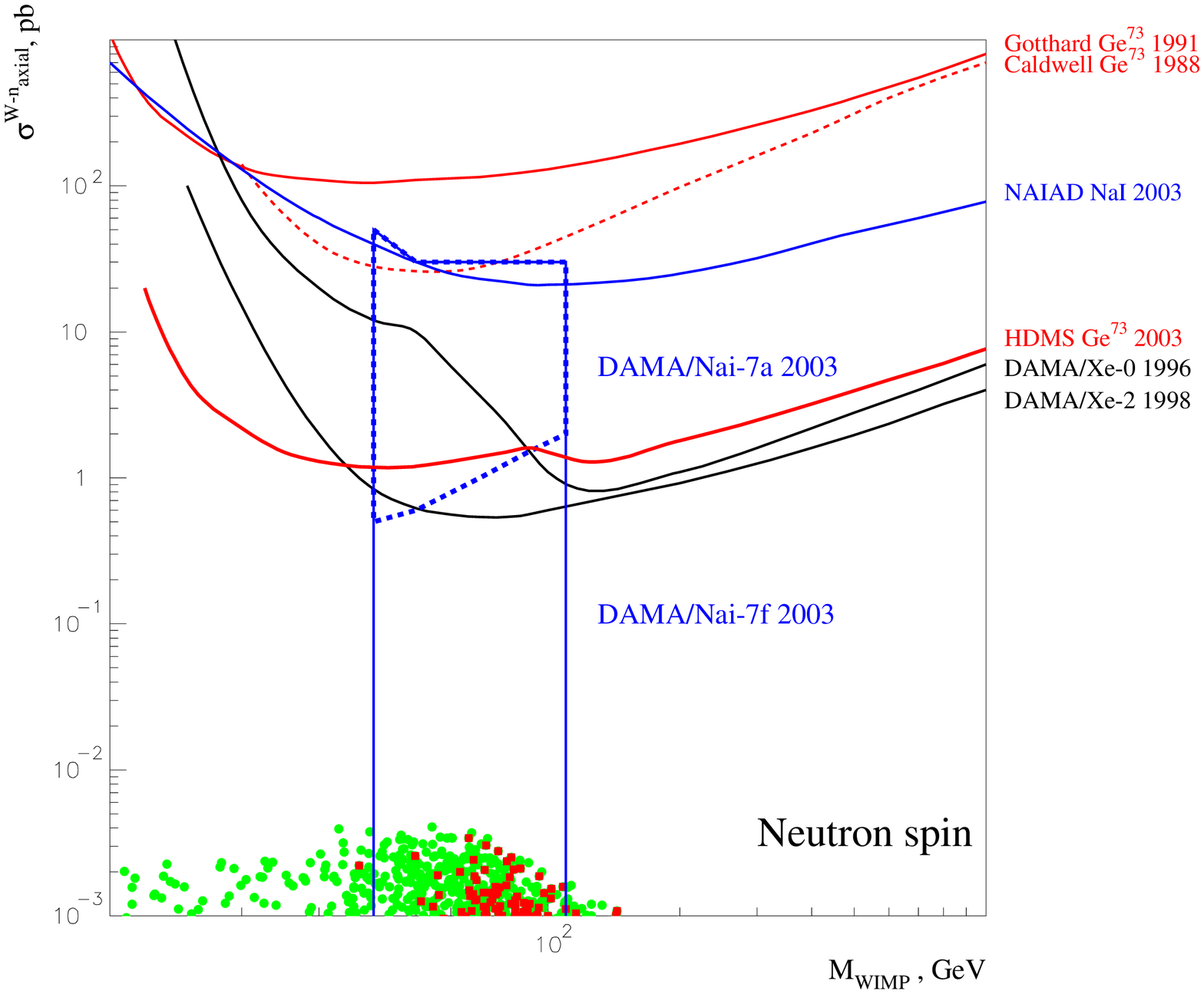}}
\end{picture}
\caption{Exclusion curves for the spin-dependent WIMP-neutron cross sections
        ($\sigma^{n}_{{\rm SD}}$ versus the WIMP mass).
        DAMA/NaI-7a(f) contours for the WIMP-neutron SD interaction
        (subdominating in $^{127}$I) 
        are obtained from the relevant figures of   
\cite{Bernabei:2003za,Bernabei:2001ve}. 
        Note that the NAIAD curve here corresponds to the 
        WIMP-neutron SD interaction subdominant for $^{127}$I. 
        The WIMP-proton SD interaction dominates for this nucleus.
        The curve was obtained in the approach of 
\cite{Tovey:2000mm}. 
        It is much weaker in comparison with the both DAMA/Xe and HDMS curves. 
From \cite{Bednyakov:2004be}.}
\label{Spin-n} 
\end{figure}

        One can see from 
(\ref{Definitions.cross.section})--(\ref{a_pn})
        that the direct dark matter search 
        experiments supply us with only three
        different constants for the 
        underlying SUSY theory from 
        non-observation of a DM signal ($c_0$, $a_p$ and $a_n$, or 
$\sigma^{p}_{{\rm SI}}(0)$, $\sigma^{p}_{{\rm SD}}(0)$ and  
$\sigma^{n}_{{\rm SD}}(0)$), 
        provided the DM particle is the lightest SUSY particle (LSP) 
        neutralino
\cite{Bednyakov:1994te}.
\begin{figure}[ht] 
\begin{picture}(60,90) 
\put(-3,-42){\includegraphics{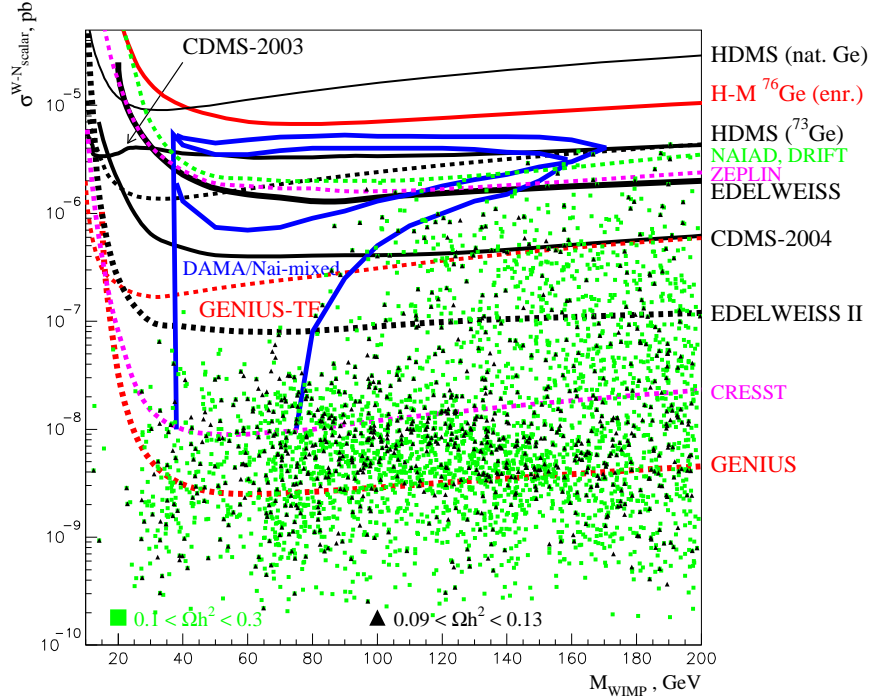}}
\end{picture}
\caption{WIMP-nucleon cross section 
        limits in pb for spin-independent (scalar) interactions as 
        a function of the WIMP mass in GeV. 
        Shown are contour lines for some of the present experimental limits 
        (solid lines) and some of projected experiments (dashed lines). 
        All curves are obtained in 
        the one-coupling dominance approach with $\sigma^{}_{\rm SD}=0$.
        For example, 
        the closed DAMA/NaI contour corresponds to complete neglection 
        of SD WIMP-nucleon interaction.
        Only the open DAMA contour is obtained in 
\cite{Bernabei:2003za} 
        with the assumption that $\sigma^{}_{\rm SD}=0.08$ pb $> 0$.
From \cite{Bednyakov:2004be}.
        }
\label{Scalar-2004} 
\end{figure} 
        These constraints are traditionally presented
        in the form of sets of exclusion curves for the
        spin-independent (scalar) nucleon-WIMP 
(Fig.~\ref{Scalar-2004}), spin-dependent (axial) proton-WIMP
(Fig.~\ref{Spin-p}) and spin-dependent neutron-WIMP cross sections
(Fig.~\ref{Spin-n}) as functions of the WIMP mass.
        From 
(\ref{Definitions.spin.zero.momentum})
        one can also see that contrary to the SI case
(\ref{Definitions.scalar.zero.momentum}) 
        both proton $\langle{\bf S}^A_{p}\rangle$
        and neutron $\langle{\bf S}^A_{n}\rangle$
        spin contributions simultaneously enter into formula 
(\ref{Definitions.spin.zero.momentum})
        for the SD WIMP-nucleus cross section $\sigma^A_{\rm SD}(0)$.
        Nevertheless, for 
        the most interesting isotopes either $\langle{\bf S}^A_{p}\rangle$ 
        or $\langle{\bf S}^A_{n}\rangle$ dominates
        ($\langle{\bf S}^A_{n(p)}\rangle \ll \langle{\bf S}^A_{p(n)}\rangle$)
\cite{Bednyakov:2004be,Bednyakov:2004xq}.

        In earlier considerations 
\cite{Smith:1990kw,Engel:1992bf,Ellis:1988sh,%
Goodman:1985dc,Drukier:1986tm,Engel:1989ix}
        one reasonably assumed that the nuclear spin was carried by the ``odd''
        unpaired group of protons or neutrons and only one of either 
        $\langle{\bf S}^A_n\rangle$ or $\langle{\bf S}^A_p\rangle$ was 
        non-zero.
        In this case all possible non-zero-spin target nuclei can 
        be classified into n-odd and p-odd groups.
        Following this classification,
        the current experimental situation
        for the spin-dependent WIMP-{\bf proton} 
        and WIMP-{\bf neutron} cross sections are 
        naturally presented separately in 
Fig.~\ref{Spin-p} and
Fig.~\ref{Spin-n}.
        The DAMA/NaI-7 contours for the WIMP-proton SD interaction 
        (dominating in $^{127}$I) obtained on the basis of the positive 
        signature of the annual modulation (closed contour)
\cite{Bernabei:2003za} and within the mixed coupling framework (open contour)
\cite{Bernabei:2001ve} are also presented in 
Fig.~\ref{Spin-p}. 
        Similarly, the DAMA/NaI-7 
\cite{Bernabei:2003za} contours for the WIMP-neutron SD interaction
        (subdominant in $^{127}$I) are given in
Fig.~\ref{Spin-n}. 
        One can also expect some 
        exclusion curves for the SD cross section from the CDMS 
\cite{Akerib:2004fq} and EDELWEISS 
\cite{Sanglard:2004kb} experiments
        with natural-germanium bolometric detectors
        (due to a small Ge-73 admixture). 
        The scatter plots for the SD LSP-proton and LSP-neutron
        cross sections calculated in the effMSSM from 
\cite{Bednyakov:2004be} are also given in 
Figs.~\ref{Scalar-2004}--\ref{Spin-n}.

        We would like to stress that the calculated scatter 
        plots for $\sigma^p_{\rm SD}$ 
(Fig.~\ref{Spin-p})
        are obtained without any assumption 
        of $\sigma^n_{\rm SD}=0$ (and $\sigma^p_{\rm SI}=0$), 
        but the experimental exclusion curves for $\sigma^p_{\rm SD}$ 
        were traditionally extracted from the data 
        with the spin-neutron (and scalar) contribution
        fully neglected,
        i.e. under the assumption that 
        $\sigma^n_{\rm SD}=0$ (and $\sigma^p_{\rm SI}=0$).
        This 
{\bf one-spin-coupling dominance scheme} 
        (always used before new approaches were proposed in
\cite{Tovey:2000mm} and in 
\cite{Bernabei:2003za,Bernabei:2000qi,Bernabei:2003wy}) 
        gave a bit too pessimistic exclusion curves,
        but allowed direct comparison of sensitivities 
        for different experiments.
        More stringent constraints on $\sigma^p_{\rm SD}$ can be obtained 
\cite{Tovey:2000mm,Bernabei:2003za,Bernabei:2000qi,Bernabei:2003wy}
        by assuming both 
        $\sigma^p_{\rm SD}\neq 0$ and $\sigma^n_{\rm SD}\neq 0$
        (although the contribution of the neutron spin is
        usually very small because
        $\langle{\bf S}^A_{n}\rangle \ll \langle{\bf S}^A_{p}\rangle$).
        Therefore a direct 
        comparison of the old-fashioned exclusion curves 
        with the new ones could in principle
        be misleading. 

        The same conclusion on the one-coupling dominance approach  
        to a great extent concerns 
\cite{Bernabei:2003za,Bernabei:2003wy}
        the direct comparison of the {\em old}\/ SI exclusion curves 
        (obtained with zero SD contribution, $\sigma^{}_{\rm SD}=0$)
        with the {\em new}\/ SI exclusion curves 
        (obtained with non-zero SD contribution, 
        $\sigma^{}_{\rm SD}>0$)
        as well as with the results of the SUSY calculations.
        One can see from 
Fig.~\ref{Scalar-2004} that the {\em new-type}\/ DAMA/NaI open contour
        (when $\sigma^{}_{\rm SD}>0$)
        is in agreement with the best exclusion curves of the  
        CDMS and EDELWEISS as well as with SUSY calculations.
        One knows that both these experiments have natural
        germanium (almost pure spinless) 
        as a target and therefore have no sensitivity to the 
        spin-dependent WIMP-nucleon couplings 
        (for them $\sigma^{}_{\rm SD}\equiv 0$). 
        Therefore, these experiments exclude only the pure 
        SI interpretation of the DAMA annual modulation signal 
\cite{Akerib:2004fq,Chardin:2004ry,Kurylov:2003ra,Copi:2000tv,Copi:2002hm}.
        The statement that this DAMA result 
        is {\em completely}\ excluded by the results of these 
        cryogenic experiments and 
        is inconsistent with the SUSY interpretation (see, for example,
\cite{Drees:2004db}) 
        is simply wrong
        (see also discussions in 
\cite{Kurylov:2003ra,Gelmini:2004gm}).

        The event-by-event CDMS and EDELWEISS background discrimination
        (via simultaneous charge and phonon signal measurements) is 
        certainly very important. 
        Nevertheless the DAMA annual signal modulation 
        is one of a few available 
        {\em positive}\/ signatures of 
        WIMP-nucleus interactions and the importance 
        of its observation goes far beyond a simple 
        background reduction. 
        Therefore, to completely exclude the DAMA result, a new experiment, 
        being indeed sensitive to the modulation signal, 
        would have to exclude this 
        modulation signal on the basis of 
        the same or much better statistics. 

         Furthermore, taking seriously 
         the positive DAMA result together with 
         the negative results of the CDMS and EDELWEISS
         as well as the results of
\cite{Savage:2004fn}
         one can arrive at a conclusion about simultaneous
         existence and 
         importance of both SD and SI WIMP-nucleus interactions.

\section{Mixed spin-scalar WIMP-nucleon couplings}
        More accurate calculations of spin nuclear structure 
(see a review in
\cite{Bednyakov:2004xq})
        demonstrate that contrary to the simplified odd-group approach both
        $\langle{\bf S}^A_{p}\rangle$ and $\langle{\bf S}^A_{n}\rangle$ 
        differ from zero, but nevertheless 
        one of these spin quantities always dominates
        ($\langle{\bf S}^A_{p}\rangle \ll \langle{\bf S}^A_{n}\rangle$, or
         $\langle{\bf S}^A_{n}\rangle \ll \langle{\bf S}^A_{p}\rangle$). 
{\it If together}\ with the dominance like 
        $\langle{\bf S}^A_{p(n)}\rangle \ll \langle{\bf S}^A_{n(p)}\rangle$
        one would have WIMP-proton and WIMP-neutron couplings
        of the same order of magnitude
        ({\it not} $a_{n(p)}\! \ll\! a_{p(n)}$), 
        the situation could look like that in the odd-group model
        and one could safely (at the current level of accuracy)
        neglect subdominant spin contribution in the data analysis.
        Indeed,  
        very large or very small ratios $\sigma_p/\sigma_n
        \sim a_{p}/a_{n}$ would correspond to the neutralinos which are
        extremely pure gauginos. 
        In this case $Z$-boson exchange in the SD interactions is absent 
        and only sfermions make contributions to the SD cross sections. 
        This is a very particular case
        which is also currently in disagreement with the experiments.
        We have checked the relation  
        $|a_n|/|a_p| \approx O(1)$ for large LSP masses in
\cite{Bednyakov:2003wf}. 
        For relatively low LSP masses $m_\chi < 200$~GeV in effMSSM 
\cite{Bednyakov:1999vh}--\cite{Bottino:2000jx}
        the $a_n$-to-$a_p$ ratio is located within the bounds
\cite{Bednyakov:2004be}:
\begin{equation}
\label{an2ap-ratio-bounds}
0.5 < \left|\frac{a_n}{a_p} \right| <  0.8.
\end{equation}
        Therefore the couplings are almost the same
        and one can quite safely use the {\em clear}\/ 
        ``old'' n-odd and p-odd group
        classification of non-zero-spin targets and neglect, for example, 
        the $\langle{\bf S}^A_{p}\rangle$-spin 
        contribution in the analysis of the 
        DM data for a nuclear target with 
        $\langle{\bf S}^A_{p}\rangle \ll \langle{\bf S}^A_{n}\rangle$.
        Furthermore, when one compares in the same figure
        the exclusion curve for SD WIMP-proton coupling
        obtained without the subdominant SD WIMP-neutron contribution 
        (all curves in 
Fig.~\ref{Spin-p} except the NAIAD one 
\cite{Ahmed:2003su} and the Tokyo-LiF one 
\cite{Miuchi:2002zp})
        with the curve from the approach of 
\cite{Tovey:2000mm}, when the subdominant contribution is included
        (the NAIAD and Tokyo-LiF curves in 
Fig.~\ref{Spin-p}),
        one {\it ``artificially''}\ improves the sensitivity 
        of the {\it latter}\ curves 
        (NAIAD or Tokyo-LiF) in comparison with the former ones.
        For the sake of consistency and reliable comparisons, 
        one should coherently recalculate
        all previous curves in the new manner
\cite{Bernabei:2003za}. 

        We note that it looks like the SI contribution is 
        completely ignored in the SIMPLE experiment 
\cite{Giuliani:2003nf,Giuliani:2004uk} 
        and the DM search with NaF bolometers   
\cite{Takeda:2003km}. 
        Although $^{19}$F has the best properties 
        for investigation of WIMP-nucleon spin-dependent 
        interactions  (see, for example,
\cite{Divari:2000dc})
        it is not obvious that one should completely ignore
        spin-independent WIMP coupling with fluorine. 
        For example, in the relation  
$\sigma^A \sim \sigma^{A,p}_{\rm SD}
        \left[\frac{\sigma^A_{\rm SI}}{\sigma^{A,p}_{\rm SD}}
        +\left(1 + \sqrt{\frac{\sigma^{A,n}_{\rm SD}}{\sigma^{A,p}_{\rm SD}}}
        \right)^2 \right]$,     
         which follows from 
(\ref{Definitions.scalar.zero.momentum})--%
(\ref{Definitions.spin.zero.momentum.Bernabei}),
        it is not a priori clear that
        $\frac{\sigma^A_{\rm SI}}{\sigma^{A,p}_{\rm SD}} \ll
        \frac{\sigma^{A,n}_{\rm SD}}{\sigma^{A,p}_{\rm SD}}$,\ 
        i.e. the SI WIMP-nucleus interaction is much weaker than
        the subdominant SD WIMP-nucleus one.
        At least for isotopes with an atomic number $A>50$
\cite{Jungman:1996df,Bednyakov:1994qa}
        to neglect the SI contribution would be a larger 
        mistake than to neglect the  
        subdominant SD WIMP-neutron contribution,  
        when the SD WIMP-proton interaction dominates, 
        at the current level of sensitivity of DM experiments
\cite{Bednyakov:1999vh,Bednyakov:2002mb}. 
        From measurements with $^{73}$Ge one can extract, following 
\cite{Tovey:2000mm},
        not only the dominant constraint for the WIMP-nucleon coupling
        $a_n$ (or $\sigma_{\rm SD}^{n}$) 
        but also the constraint for the subdominant WIMP-proton coupling
        $a_p$ (or $\sigma_{\rm SD}^{p}$). 
        Nevertheless, the latter constraint will be much weaker
        in comparison with the constraints from p-odd 
        target nuclei, like $^{19}$F or $^{127}$I.  
        This fact is illustrated by the ``weak'' NAIAD (NaI, 2003) curve in 
Fig.~\ref{Spin-n}, which corresponds to the subdominant
        WIMP-neutron spin contribution 
        extracted from the p-odd nucleus $^{127}$I. 

        Therefore we would like to note that 
        the ``old'' odd-group-based approach to 
        analysis of the SD data from experiments with heavy enough
        targets (for example, Ge-73) is still quite suitable, 
        especially when it is not obvious that 
        (both) spin couplings dominate over the scalar one.

        The approach of Bernabei et al.
\cite{Bernabei:2003za,Bernabei:2001ve}
        looks more appropriate for the mixed spin-scalar 
        coupling data presentation,
        and is based on introduction of the effective 
        SD nucleon cross section $\sigma^{}_{\rm SD}(0)$
        and the coupling mixing angle $\theta$
(Eq. (\ref{effectiveSD-cs}))
        instead of $\sigma^{p}_{\rm SD}(0)$ and
        $\sigma^{n}_{\rm SD}(0)$.
        With these definitions the SD WIMP-proton and
        WIMP-neutron cross sections have the form 
$\sigma^p_{\rm SD}=\sigma^{}_{\rm SD} \cdot \cos^2 \theta$ and 
$\sigma^n_{\rm SD}=\sigma^{}_{\rm SD} \cdot \sin^2 \theta$.

\begin{figure}[p] 
\begin{picture}(100,160)
\put(  0,78){\includegraphics{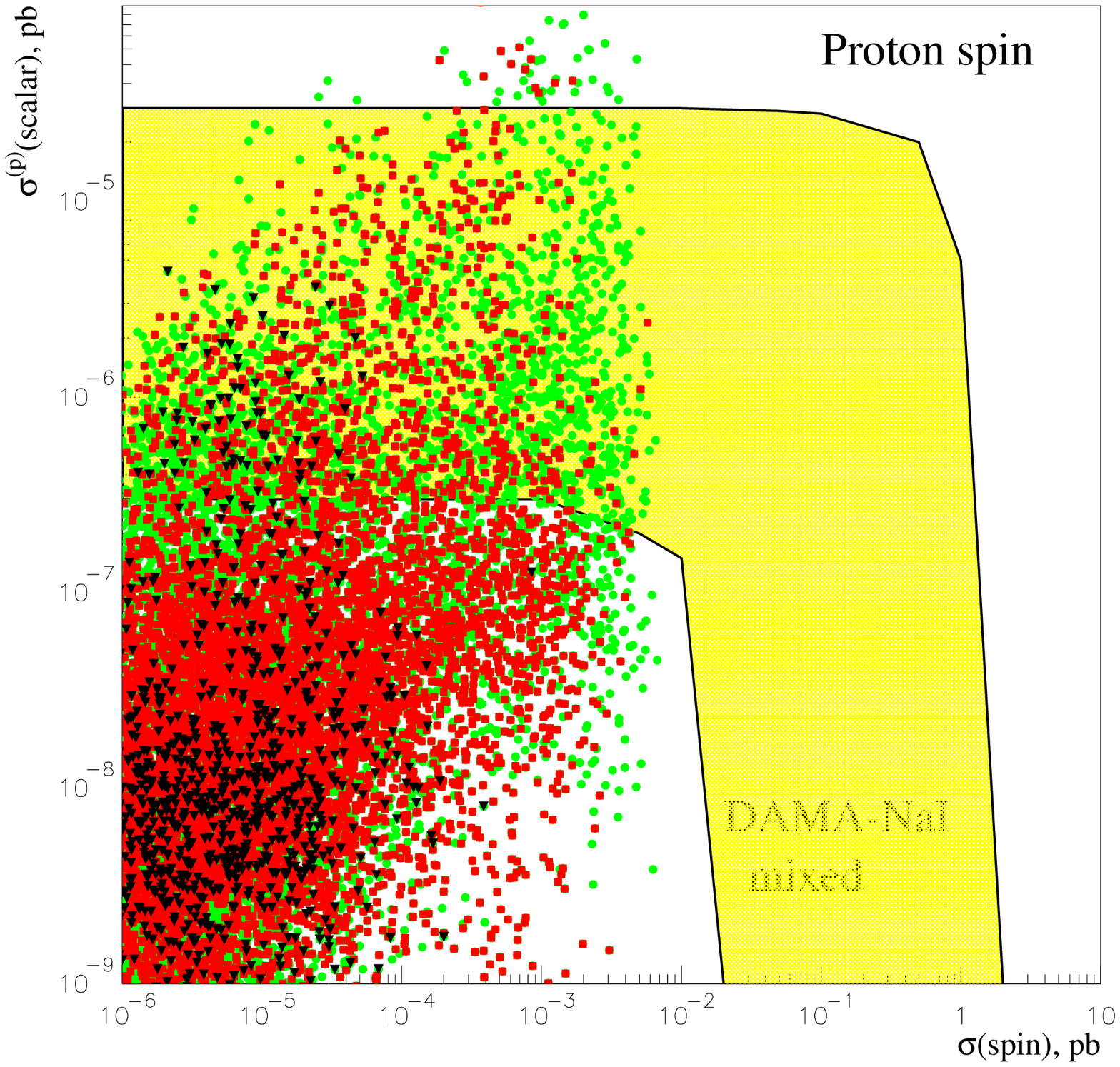}}
\put( 36,-4){\includegraphics{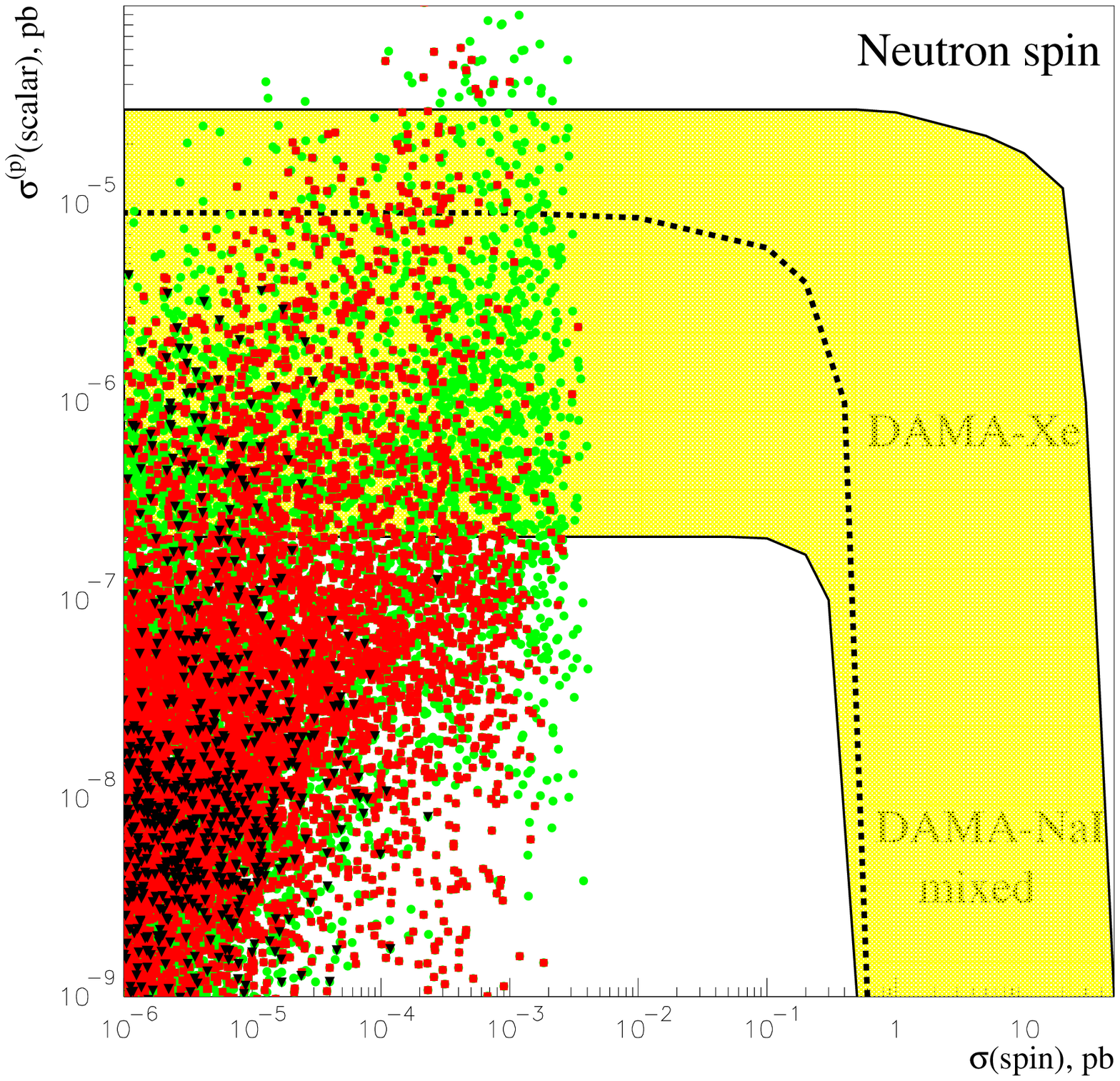}}
\end{picture}
\caption{The DAMA-NaI allowed region 
        from the WIMP annual modulation signature 
        in the ($\xi \sigma_{\rm SI}$, $\xi \sigma_{\rm SD}$) space 
        for $40<m^{}_{\rm WIMP}<110$~GeV
\cite{Bernabei:2003za,Bernabei:2001ve}.
        The left panel corresponds to the dominating (in $^{127}$I)  
        SD WIMP-proton coupling alone ($\theta$ = 0)  
        and the right panel corresponds to 
        the subdominating SD WIMP-neutron coupling alone 
        ($\theta$ = $\pi/2$). 
        The scatter plots give correlations
        between $\sigma^{p}_{{\rm SI}}$ and 
        $\sigma^{}_{{\rm SD}}$ in the effMSSM ($\xi=1$ is assumed) 
        for $m_\chi<200$~GeV 
\cite{Bednyakov:2004be}.
        In the right panel the DAMA liquid xenon exclusion curve from 
\cite{Bernabei:2001ve} is given (dashed line).
From \cite{Bednyakov:2004be}.}
\label{Bernabei:2001ve:fig}
\end{figure} 

In Fig.~\ref{Bernabei:2001ve:fig}
        the WIMP-nucleon spin and scalar mixed couplings 
        allowed by the annual modulation signature from 
        the 100-kg DAMA/NaI experiment are shown inside the shaded regions. 
        The regions from 
\cite{Bernabei:2003za,Bernabei:2001ve} 
        in the ($\xi \sigma_{\rm SI}$, $\xi \sigma_{\rm SD}$) space 
        for 40~GeV$<m^{}_{\rm WIMP}<$110~GeV cover
        spin-scalar mixing coupling for the proton ($\theta=0$ case of 
\cite{Bernabei:2003za,Bernabei:2001ve}, left panel) and 
        spin-scalar mixing coupling for the neutron 
        ($\theta$ = $\pi/2$, right panel). 
        From nuclear physics one has for the proton spin dominated
        $^{23}$Na and $^{127}$I \
        $\frac{\langle {\bf S}_n\rangle}{\langle {\bf S}_p\rangle}< 0.1$ 
        and 
        $\frac{\langle {\bf S}_n\rangle}
        {\langle {\bf S}_p\rangle}< 0.02 \div 0.23$,
        respectively.
        For $\theta=0$ 
        the DAMA WIMP-proton spin constraint is the severest one
        due to the p-oddness of the I target
(see Fig.~\ref{Spin-p}).  

        In the right panel of  
Fig.~\ref{Bernabei:2001ve:fig}
        we present the exclusion curve (dashed line) for the
        WIMP-neutron 
        spin coupling from the odd-neutron  
        isotope $^{129}$Xe obtained under the mixed coupling assumptions  
\cite{Bernabei:2001ve} from the DAMA-LiXe (1998) experiment 
\cite{Bernabei:2002qg,Bernabei:1998ad,Bernabei:2002af}.
        For the DAMA NaI detector the 
        $\theta=\pi/2$ means no 
${\langle {\bf S}_p\rangle}$ contribution at all. 
        Therefore, in this case DAMA gives the 
        subdominant ${\langle {\bf S}_n\rangle}$ 
        contribution alone,
        which could be compared further with the dominant 
        ${\langle {\bf S}_n\rangle}$ contribution in $^{73}$Ge.
        
        The scatter plots in
Fig.~\ref{Bernabei:2001ve:fig} give 
        $\sigma^{p}_{{\rm SI}}$ as a function of 
        $\sigma^p_{{\rm SD}}$ (left panel) and  
        $\sigma^n_{{\rm SD}}$  (right panel)
        calculated in the effMSSM 
\cite{Bednyakov:2004be}.
        Filled circles (green) correspond to the relic neutralino density 
        $0.0< \Omega_\chi h^2_0<1.0$,
        squares (red) correspond to the subdominant relic neutralino
        contribution 
        $0.002 < \Omega_\chi h^2_0<0.1$         
        and triangles (black)
        correspond to the 
        WMAP density constraint 
        $0.094 < \Omega_\chi h^2_0<0.129$
\cite{Spergel:2003cb,Bennett:2003bz}. 

        The constraints on the SUSY parameter space 
        within the mixed coupling 
        framework in Fig.~\ref{Bernabei:2001ve:fig} 
        are in general 
        much stronger in comparison with the 
        traditional approach based on the one-coupling dominance
(Figs.~\ref{Scalar-2004}, \ref{Spin-p} and \ref{Spin-n}). 

        It follows from 
Fig.~\ref{Bernabei:2001ve:fig} 
        that when the LSP is the subdominant DM particle (squares in the
        figure), SD WIMP-proton and WIMP-neutron cross sections 
        at a level of $3\div5\cdot 10^{-3}$~pb are allowed,
        but the WMAP relic density constraint (triangles)
        together with the DAMA restrictions leaves only 
        $\sigma_{\rm SD}^{p,n}<3\cdot 10^{-5}$~pb
        without any visible reduction of allowed values for
        $\sigma^p_{\rm SI}$.
        In general, according to the DAMA restrictions, 
        very small SI cross sections are completely excluded, 
        only $\sigma^p_{\rm SI}> 3\div5\cdot 10^{-7}$~pb are allowed. 
        As to the SD cross section, the situation is not clear,
        because for the allowed values of the SI contribution  
        the SD DAMA sensitivity did not yet reach  
        the calculated upper bound for the SD LSP-proton
        cross section of $5\cdot 10^{-2}$~pb 
        (for the current nucleon spin structure from  
\cite{Ellis:2000ds}).

\section{The mixed couplings case for high-spin \boldmath $^{73}$Ge}
        There are many measurements with p-odd nuclei
        and there is a lack of data for n-odd nuclei, i.e. 
        for $\sigma_{\rm SD}^{n}$. 
        From our point of view this lack of $\sigma_{\rm SD}^{n}$ 
        measurements can be 
        filled with new data expected from the HDMS experiment with 
        the high-spin isotope $^{73}$Ge
\cite{Klapdor-Kleingrothaus:2002pi}.
        This isotope looks with a good accuracy 
        like an almost pure n-odd group nucleus with 
        $\langle {\bf S}_{n}\rangle\! \gg\! \langle {\bf S}_{p}\rangle$
(Table~\ref{Nuclear.spin.main.table.71-95}).
        The variation in  $\langle {\bf S}_{p}\rangle$
        and $\langle {\bf S}_{n}\rangle$
        in the table reflects the level of  
        inaccuracy and complexity 
        of the current nuclear structure calculations.
\begin{table}[h!] 
\caption{All available calculations in different nuclear models for the
        zero-momentum spin structure (and predicted magnetic moments $\mu$) 
        of the $^{73}$Ge nucleus. 
        The experimental value of the magnetic moment given in the brackets 
        is used as input in the calculations.
\label{Nuclear.spin.main.table.71-95}}
\begin{center}
\small\smallskip
\begin{tabular}{lrrr}
\hline\hline
$^{73}$Ge~($G_{9/2}$) & ~~~~~$\langle {\bf S}_p \rangle$ & 
~~~~~$\langle {\bf S}_n \rangle$ & ~~~~~$\mu$ (in $\mu_N$) \\ \hline
ISPSM, Ellis--Flores~\cite{Ellis:1988sh,Ellis:1991ef}
        &    0    & $0.5$               & $-1.913$ \\ 
OGM, Engel--Vogel~\cite{Engel:1989ix}   
        &    0    & $0.23$      &$(-0.879)_{\rm exp}$ \\ 
IBFM, Iachello at al.~\cite{Iachello:1991ut} and \cite{Ressell:1993qm}
        &$-0.009$ & $0.469$ &$-1.785$\\ 
IBFM (quenched), 
        Iachello at al.~\cite{Iachello:1991ut} and \cite{Ressell:1993qm}
        &$-0.005$  & $0.245$ &$(-0.879)_{\rm exp}$ \\
TFFS, Nikolaev--Klapdor-Kleingrothaus, \cite{Nikolaev:1993dd} 
        &$0$   & $0.34$ & --- \\ 
SM (small), Ressell at al.~\cite{Ressell:1993qm} 
        &$0.005$   & $0.496$ &$-1.468$ \\ 
SM (large), Ressell at al.~\cite{Ressell:1993qm} 
        &$0.011$   & $0.468$ &$-1.239$ \\ 
SM (large, quenched), Ressell at al.~\cite{Ressell:1993qm} 
        &$0.009$   & $0.372$ &$(-0.879)_{\rm exp}$ \\ 
``Hybrid'' SM, Dimitrov at al.~\cite{Dimitrov:1995gc}           
        & $0.030$ & $0.378$ & $-0.920$ \\ 
\hline\hline
\end{tabular} \end{center}
\end{table} 

        In the mixed spin-scalar coupling case 
        the direct detection rate 
(\ref{Definitions.diff.rate}) in $^{73}$Ge
        integrated over recoil energy from the threshold energy, $\eth$, 
        to the maximal energy, $\emx$, 
        is a sum of the SD and SI contributions: 
\begin{eqnarray}
\label{for-toy-mixig}\label{for-Ge-73}
R(\eth, \emx)&=&
        \alpha(\eth,\emx,m_\chi)\,\sigma^p_\SI
        +\beta(\eth,\emx,m_\chi)\,\sigma^{}_\SD;\\
&&\alpha(\eth,\emx,m_\chi)
        = N_T \frac{\rho_\chi M_A}
                {2 m_\chi \mu_p^2 } A^2 
        A_\SI(\eth,\emx),\nonumber \\
&&\beta(\eth,\emx,m_\chi)
        =       N_T \frac{\rho_\chi M_A}
                 {2 m_\chi \mu_p^2 } 
        \frac43 \frac{J+1}{J}
        \left( \langle {\bf S}^A_p\rangle \cos\theta
              +\langle {\bf S}^A_n\rangle \sin\theta   
       \right)^2
        A_\SD(\eth,\emx); \nonumber \\
&& A_{\SI,\SD}(\eth,\emx) = 
        \frac{\langle {v}\rangle}
         {\langle {v}^2 \rangle}
        \int_\eth^\emx d\ER F^2_{\SI,\SD}(\ER)I(\ER).
\label{for-toy-mixig-FF}
\end{eqnarray}
        To estimate the event rate 
(\ref{for-toy-mixig})   
        one should know a number of quite uncertain 
        astrophysical and nuclear structure parameters
        as well as the precise characteristics of the experimental setup
        (see, for example, the discussions in 
\cite{Bernabei:2003za,Bernabei:2003xg}).
\begin{figure}[th] 
\begin{picture}(100,70)
\put(17,-3){\includegraphics{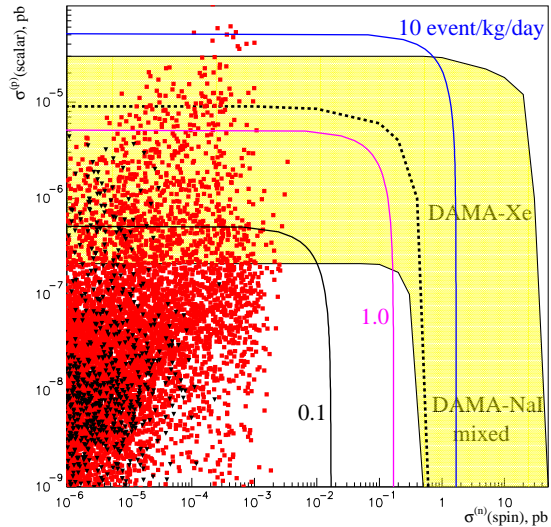}}
\end{picture}
\caption{ 
        Solid lines (marked with numbers of $R(15,50)$ in events/kg/day)
        show the sensitivities of the HDMS setup with $^{73}$Ge
        within the framework of mixed SD WIMP-neutron 
        and SI WIMP-nucleon couplings.
        The DAMA-NaI region for the subdominant SD WIMP-neutron coupling 
        ($\theta$ = $\pi/2$) is from 
Fig.~\ref{Bernabei:2001ve:fig}.
        Scatter plots give correlations
        between $\sigma^{p}_{{\rm SI}}$ and 
        $\sigma^{n}_{{\rm SD}}$ in the effMSSM 
        for $m_\chi<200$~GeV
\cite{Bednyakov:2004be}.
        Squares (red) correspond to the subdominant relic neutralino
        contribution $0.002 < \Omega_\chi h^2_0<0.1$    
        and triangles (black)
        correspond to the WMAP relic neutralino density 
        $0.094 < \Omega_\chi h^2_0<0.129$. 
        The dashed line from 
\cite{Bernabei:2001ve}
        shows the DAMA-LiXe (1998) exclusion curve 
        for $m^{}_{\rm WIMP}=50$~GeV. 
From \cite{Bednyakov:2004be}.}
\label{hdms-toy}
\end{figure} 

        We neglect 
        the subdominant contribution from 
        the WIMP-proton spin coupling proportional 
        to $\langle {\bf S}^A_p\rangle$ for $^{73}${Ge}.
        We consider only a simple spherically symmetric isothermal 
        WIMP velocity distribution 
\cite{Drukier:1986tm,Freese:1988wu}
        and do not go into details of any possible
        and in principle important 
        uncertainties (and/or modulation effects) 
        of the Galactic halo WIMP distribution 
\cite{Kinkhabwala:1998zj}--\cite{Vergados:2000cp}.
        For simplicity we use the Gaussian 
        scalar and spin nuclear form-factors from 
\cite{Ellis:1991ef,Ellis:1993ka}.
        With formulas 
(\ref{for-Ge-73}) we performed a simple estimation of prospects
        for the DM search and SUSY constraints with the high-spin $^{73}$Ge
        detector HDMS 
        taking into account the 
        available results from the 
        DAMA-NaI and LiXe experiments
\cite{Bernabei:2003za,Bernabei:2000qi,Bernabei:2003wy,%
      Bernabei:2002qg,Bernabei:1998ad,Bernabei:2002af}.

        The Heidelberg Dark Matter Search (HDMS) experiment
        uses a special configuration of 
        two Ge detectors to efficiently reduce the background
\cite{Klapdor-Kleingrothaus:2000uh}.
        From the first preliminary
        results of the HDMS experiment 
        with the inner HPGe crystal of enriched $^{73}$Ge
\cite{Klapdor-Kleingrothaus:2002pi}
        we can estimate the current background event rate 
        $R(\eth, \emx)$ integrated here from  
        the ``threshold'' energy $\eth=15$~keV to the ``maximal'' 
        energy $\emx=50$~keV. 
        We obtain $R(15,50)\approx 10$ events/kg/day.
        A substantial improvement of the background 
        (up to an order of magnitude) is further expected for
        the setup in the Gran Sasso Underground Laboratory.
In Fig.~\ref{hdms-toy}
        solid lines for the integrated rate $R(15,50)$
        marked with numbers 10, 1.0 and 0.1 (in events/kg/day) 
        present our exclusion curves for $m^{}_{\rm WIMP}=70$~GeV
        expected from the HDMS setup with $^{73}$Ge
        within the framework of the mixed SD WIMP-neutron 
        and SI WIMP-nucleon couplings.
        Unfortunately, the current background index 
        for HDMS is not yet optimized,
        and the relevant 
        exclusion curve (marked with 10 events/kg/day) 
        has almost the same strength to reduce $\sigma^{n}_{{\rm SD}}$ 
        as the dashed curve from the DAMA experiment with liquid Xe
\cite{Bernabei:2001ve} obtained for $m^{}_{\rm WIMP}=50$~GeV
        (better sensitivity is expected with HDMS for 
        $m^{}_{\rm WIMP}<40$~GeV). 
        However, both experiments lead to some sharper restriction 
        for $\sigma^{n}_{\SD}$ than obtained by DAMA (see 
Fig.~\ref{hdms-toy}). 
        An order of magnitude improvement of the 
        HDMS sensitivity (curve marked with 1.0)
        will supply us with the best exclusion curve for 
        the SD WIMP-neutron coupling, but this sensitivity is not
        yet enough to reach the calculated 
        upper bound for $\sigma^{n}_{{\rm SD}}$.   
        This sensitivity also could reduce the 
        upper bound for the SI WIMP-proton coupling $\sigma^{p}_{{\rm SI}}$
        to a level of $10^{-5}$~pb.
        Nevertheless, only an {\it additional} about-one-order-of-magnitude
        HDMS sensitivity improvement is needed to obtain
        decisive constraints on 
        $\sigma^{p}_{{\rm SI}}$ as well as on
        $\sigma^{n}_{{\rm SD}}$.
        In this case only quite narrow bounds  
        for these cross sections will be allowed
        (below the curve marked by 0.1 and above the 
        lower bound of the DAMA-NaI mixed region). 

\section{Conclusion} 
        In this paper we argue that
        potentially misleading discrepancies between the results of 
        different dark matter search experiments 
        (for example, DAMA vs CDMS and EDELWEISS) 
        as well as between the data and the SUSY calculations  
        can be avoided by using the 
        mixed spin-scalar coupling approach, where the 
        spin-independent and spin-dependent 
        WIMP-nucleon couplings are a priori considered to be
        {\em both}\/ non-zero.  
        There is generally some possible incorrectness 
        in the direct comparison of the exclusion curves 
        for the WIMP-proton(neutron) spin-dependent cross section 
        obtained with and without the
        non-zero WIMP-neutron(proton) spin-dependent contribution.

        On the other hand, nuclear spin structure calculations
        show that usually one, WIMP-proton $\langle{\bf S}^A_{p}\rangle$ 
        or WIMP-neutron $\langle{\bf S}^A_{n}\rangle$, 
        nuclear spin dominates 
        and the WIMP-proton and WIMP-neutron effective 
        couplings $a_{n}$ and $a_p$ are of the same order of magnitude.
        Therefore at the current level
        of accuracy it looks 
        safe to neglect 
        subdominant WIMP-nucleon contributions when one  
        analyses the data from non-zero-spin targets.
        The clear ``old'' odd-group-based approach to the
        analysis of the SD data from experiments with heavy enough
        targets (for example, Ge-73) is still quite suitable. 

        Furthermore the above-mentioned incorrectness concerns 
        to a great extent the direct comparison
        of spin-dependent exclusion curves obtained with and without non-zero
        spin-independent contributions
\cite{Bernabei:2003za,Bernabei:2003wy}.
        Taking into account both 
        spin couplings $a_p$ and $a_n$ but ignoring 
        the scalar coupling $c_0$, one can easily arrive at  
        a misleading conclusion 
        especially for not very light target nuclei
        when it is not obvious that 
        (both) spin couplings dominate over the scalar one.

        To be consistent,  one has 
        to use a mixed spin-scalar coupling approach
        as for the first time proposed by the DAMA collaboration
\cite{Bernabei:2000qi,Bernabei:2003za,Bernabei:2003wy}. 
        We applied the spin-scalar coupling approach to estimate 
        future prospects of the HDMS experiment with 
        the neutron-odd group high-spin isotope $^{73}$Ge. 
        Although even at the present accuracy 
        the odd-neutron nuclei $^{73}$Ge and $^{129}${Xe} 
        lead to somewhat sharper 
        restrictions for $\sigma^{n}_{\SD}$ than obtained by DAMA,  
        we found that the current accuracy of measurements with 
        $^{73}$Ge (as well as with $^{129}$Xe and NaI)
        has not yet reached a level which allows us to obtain 
        new decisive constraints on the SUSY parameters.

\smallskip
      This investigation was partly supported by the RFBR 
      (Project 02-02-04009).

\printindex
\end{document}